Machine-Learning-based Prediction of Lattice Thermal Conductivity for Half-Heusler Compounds using Atomic Information

Hidetoshi Miyazaki[1*], Tomoyuki Tamura[1, 2], Masashi Mikami[3], Kosuke Watanabe[1], Ide Naoki[1], Osman Murat Ozkendir[4], and Yoichi Nishino[1]

[1] *Nagoya Institute of Technology, Dept. of Physical Science and Engineering, 466-8555, Nagoya, Japan*
[2] *Center for Materials research by Information Integration, National Institute for Materials Science (NIMS), Tsukuba 305-0047, Japan*
[3] *National Institute of Advanced Industrial Science and Technology, 2266-98 Anagahora, Shimoshidami, Moriyama, Nagoya 463-8560*
[4] *Tarsus University, Institute of Natural Science, Dept. of Nanotechnology and Advanced Materials, 33400, Tarsus, Turkey*

*miyazaki@nitech.ac.jp

## ABSTRACT

The half-Heusler compound has drawn attention in a variety of fields as a candidate material for thermoelectric energy conversion and spintronics technology. This is because it has various electronic structures, such as semi-metals, semiconductors, and a topological insulator. When the half-Heusler compound is incorporated into the device, the control of high lattice thermal conductivity owing to high crystal symmetry is a challenge for the thermal manager of the device. The calculation for the prediction of lattice thermal conductivity, which is an important physical parameter for controlling the thermal management of the device, requires a calculation cost of several 100 times as much as the usual density functional theory calculation. Therefore, we examined whether lattice thermal conductivity prediction by machine learning was possible on the basis of only the atomic information of constituent elements for thermal conductivity calculated by the density functional theory calculation in various half-Heusler compounds. Consequently, we constructed a machine learning model, which can predict the lattice thermal conductivity with high accuracy from the information of only atomic radius and atomic mass of each site in the half-Heusler type crystal structure. Applying our results, the lattice thermal conductivity for an unknown half-Heusler compound can be immediately predicted. In the future, low-cost and short-time development of new functional materials can be realized, leading to breakthroughs in the search of novel functional materials.

## I. INTRODUCTION

Thermal conductivity is a physical quantity representing how heat is transferred from one side of a material to the other when thermal energy is applied. It is one of the most fundamental and important physical quantities. Moreover, it is an important physical quantity in terms of application, which is necessary for the understanding of thermal management to ensure the performance, life-time, and safety for thermoelectric energy conversion devices, and spintronics technology.

Thermal conductivity can be divided into electron and lattice contributions. The thermal conductivity of electrons can be determined from the electrical conductivity using the Wiedemann-Franz law. Half-Heusler compounds show high lattice thermal conductivities due to their high degrees of crystal symmetry. Therefore, it is necessary to know the exact lattice thermal conductivity for thermal management in half-Heusler devices. Theoretical predictions of the lattice thermal conductivity of solids can be made using nonequilibrium molecular dynamics simulations [1–7] or the density functional theory (DFT) calculations [8–16]. The prediction of thermal conductivity by nonequilibrium molecular dynamics requires an enormous amount of computational time because time evolution must be calculated for numerous atomic movements. However, because DFT can accurately calculate interactions between atoms, thermal conductivity can be predicted several hundred times. Presently, theoretical studies of the lattice thermal conductivity are limited to systems with a small number of atoms in the unit cell, such as simple pure metals [8, 11, 16], binary materials [9, 13–17], and full and half-Heusler compounds [12, 18–21]. Therefore, it is difficult to perform comprehensive lattice thermal conductivity calculations for a large number of materials.

Cubic half-Heusler compounds have drawn significant attention in various fields as candidate materials for thermoelectric energy conversion [22–27] and spintronics technology [28–32].This is because it has various electronic structures, such as semi-metals, semiconductors, and a topological insulator. Jesus Carrete *et al*. [33] and Jianghui Liu *et al*. [34] attempted to predict the lattice thermal conductivity of half-Heusler compounds obtained from the results of the DFT calculations by machine learning (ML). Jesus Carrete *et al*. reported that the lattice thermal conductivity of half-Heusler compounds can be predicted in the range of 10% using the Young's modulus value obtained from the DFT calculations as the descriptor for the ML. Jianghui Liu *et al*. reported that the lattice thermal conductivity of half-Heusler compounds can be predicted in the range of 10% using the atomic numbers, atomic masses, and atomic radii of the constituent atoms of half-Heusler compounds as the descriptors for ML.

The prediction of the lattice thermal conductivity must be highly accurate to predict the performance of thermoelectric materials and thermal management of electronic devices. Therefore, we have developed a ML algorithm to predict the lattice parameter and lattice thermal conductivity of half-Heusler compounds from the atomic information of their constituent atoms (atomic radius and atomic mass) only. This algorithm can predict lattice thermal conductivity values with high accuracy of less than 4% for many half-Heusler compounds. In this study, we found that the lattice thermal conductivity of the half-Heusler compounds, which is difficult to predict using the DFT calculations, can be predicted with high accuracy by ML. This report will provide a breakthrough in the development of new materials, as it can contribute to the discovery of innovative materials for next-generation thermoelectric conversion and spintronics materials, and other technologies requiring thermal management in the future.

## II. CALCULATION METHODS

### Evaluation of site selection of the half-Heusler structure and lattice parameter

Candidate half-Heusler compounds for lattice thermal conductivity calculations were sought from the Materials Project [35]. In the crystal structure of $C1_b$-type half-Heusler compounds, there are three types of atomic sites: 4*a* (0, 0, 0), 4*b* (0.5, 0, 0), and 4*c* (0.25, 0.25, 0.25) sites. The 4*a* and 4*b* sites are crystallographically interchangeable. Three structural models can be considered as one of the three constituent elements occupies the 4*c* site and the other two elements occupy the 4a and 4b sites. The lattice parameters were optimized using the DFT calculations for the three models. By comparing the total energies of the three models, the structural model with the lowest energy was determined to be the most stable structural model for the half-Heusler compound. The DFT calculations were performed using the Vienna *ab initio* Simulation Package (VASP) [36–38]. We adopted the projector augmented-wave (PAW) method [39, 40] with the generalized gradient approximation of Perdew, Burke, and Ernzerhof [41] for the exchange–correlation interactions.

### Evaluation of lattice thermal conductivity of half-Heusler compounds

The lattice thermal conductivity of half-Heusler compounds was calculated using the Phono3py code, developed by Atushi Togo *et al* [17], for various positions of atoms in a supercell made of $2 \times 2 \times 2$ primitive cells. Although the lattice thermal conductivity shows a temperature dependence, herein, we discuss the lattice thermal conductivity at 300 K. The calculated lattice thermal conductivities and lattice parameters are summarized in Table S1 in the Supplementary materials section.

### Machine Learning of thermal conductivity of half-Heusler compounds

Figure 1 shows a flowchart of the ML used to predict the lattice parameter and lattice thermal conductivity in half-Heusler compounds. The atomic radii, $r_1$, $r_2$, $r_3$, and atomic masses, $m_1$, $m_2$, $m_3$, of the elements at 4*c*, 4*a*, and 4*b* sites in the $C1_b$-type crystal structure were used as descriptors. The Python library Pymatgen, Python Materials Genomics [42], was used to obtain the elemental information. The swapped data set of the elemental information of the 4*a* and 4*b* sites was also created because the 4*a* and 4*b* sites are interchangeable. First, to build an ML model to predict the lattice parameters, the atomic radii and masses were used as descriptors. Second, to build a ML model to predict the calculated lattice thermal conductivity, the parameters generated by various combinations of atomic masses, atomic radii, and the lattice parameters were used. The parameters of the combinations of atomic radii and atomic masses used for lattice thermal conductivity prediction are listed in Table S2 in the Supplementary materials section. Finally, to find the best combination of parameters, we used the Wrapper method with a backward feature elimination to sequentially remove unimportant parameters hence build an optimal

ML model for predicting the lattice thermal conductivity. For lattice parameter and lattice thermal conductivity prediction, the multiple linear regression and boosted decision tree regression models were used as the ML models. The hyperparameters were adjusted by random sweeps to adjust the optimal hyperparameters. Python 3.6 was used for implementing the ML model. The training and test data were split between 80% and 20%. We used 5-fold cross validation on the data set to evaluate the decision coefficients using the test data for each fold, and the mean of the decision coefficients, $R^2$, was used to evaluate the ML model.

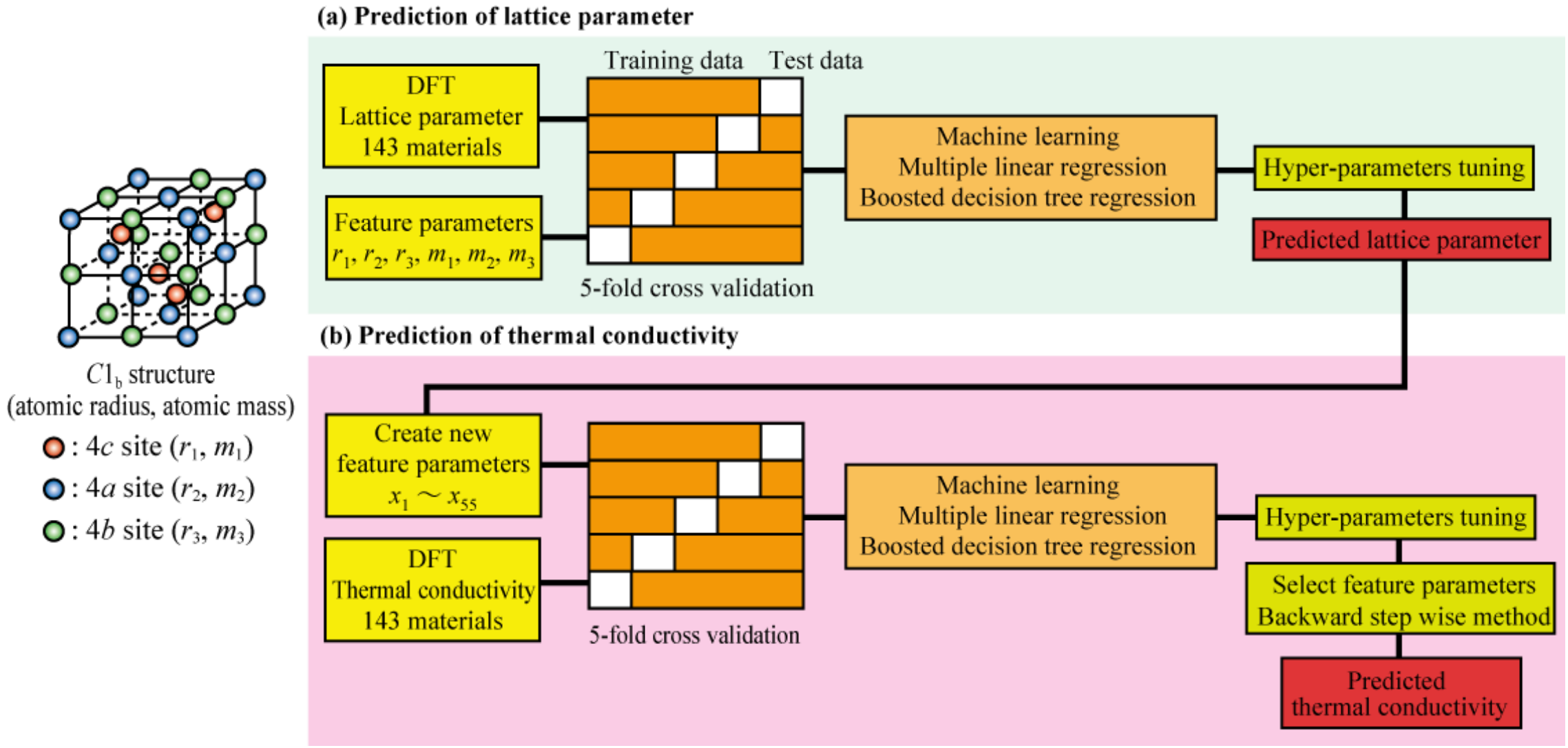

FIG. 1

Flowchart of the ML used to predict (a) lattice parameter and (b) thermal conductivity.

## III. RESULTS and DISCUSSIONS

Figures 2 (a1), (a2), and (a3) show the results of ML using multiple linear regression of the lattice parameters for various half-Heusler compounds obtained from the structural optimization by DFT calculations. ML with multiple linear regression predicts the lattice parameter with a higher accuracy using the atomic masses and atomic radii than that using the atomic radii only. As described in Table S.1, the stability of half-Heusler compounds is affected by the atomic radii of each site and the atomic masses. Therefore, besides the atomic radii, the atomic masses have an important influence on the lattice parameter determination. The prediction of the lattice parameter using multiple linear regression can be determined with an accuracy of approximately 5%, as shown in Figure 2 (a3). To determine the lattice parameter with a better accuracy, we performed ML by boosted decision tree regression as shown in Figures 2 (b1), (b2), and (b3). Using the atomic radius and mass as parameters, $R^2$ improved to 0.979. As shown in Fig. 1 (b3), the ML model reproduced the lattice parameter almost perfectly with an accuracy of approximately ± 1%. The boosted decision tree regression using the atomic radii and masses of the atoms at the three sites of the $C1_b$-type structure as a description was found to be the most suitable for predicting the lattice parameter by ML.

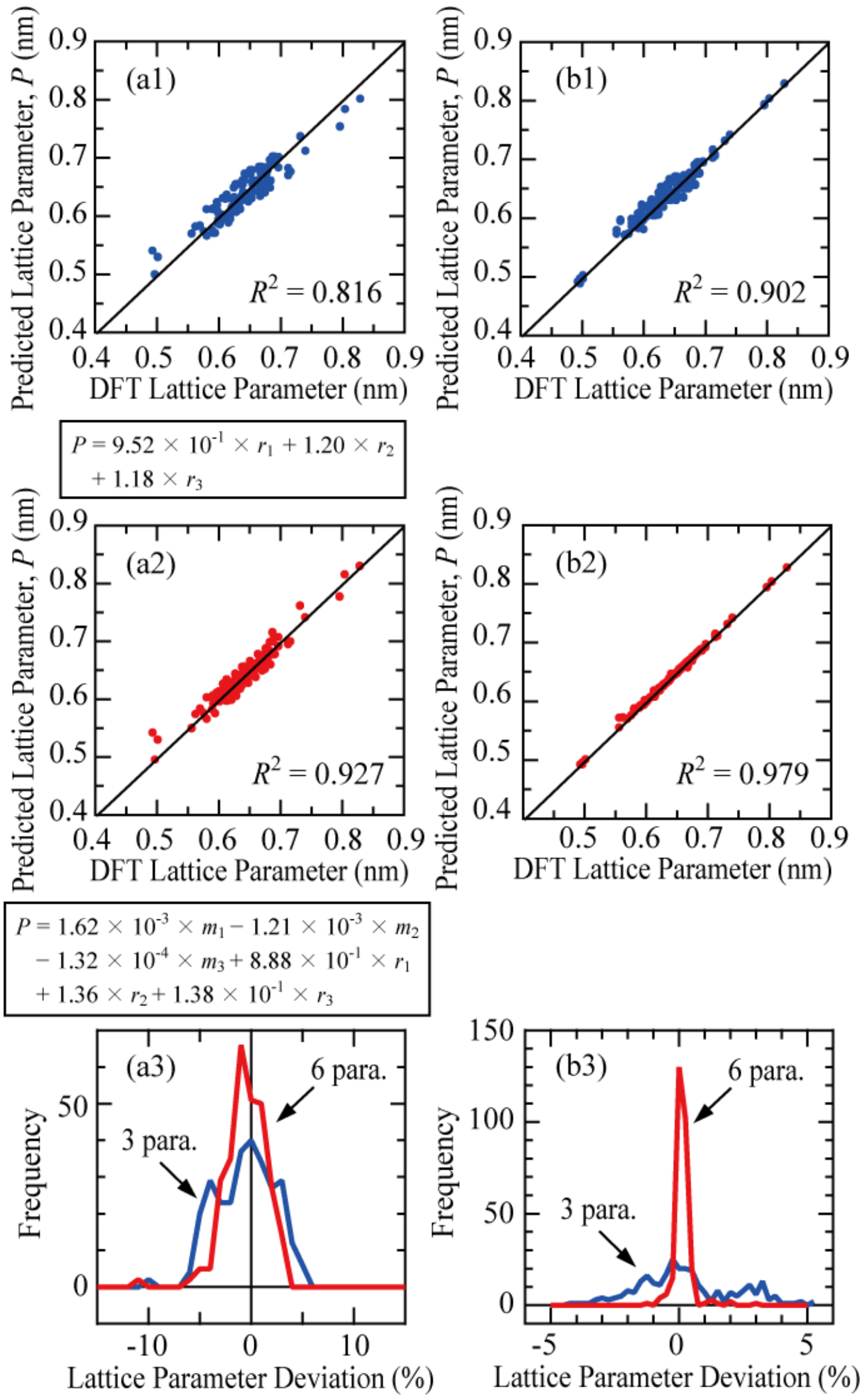

FIG. 2
Comparison between the lattice parameters predicted by the ML model of multiple linear regression (a1, a2, a3) and boosted decision tree regression (b1, b2, b3) and the lattice parameters calculated by the DFT calculation. (a1, b1) and (a2, a3) are the results of ML using a combination of atomic radii (3 parameters) and atomic mass and atomic radius (6 parameters) as descriptors, respectively. The regression equations determined by multiple linear regression are shown at the bottom of figures (a1, a2). (a3, b3) Frequency of deviations between the calculated and predicted lattice parameters.

Figures 3 show the results of the predicted lattice thermal conductivity for various half-Heusler compounds using ML with multiple linear regression and boosted decision tree regression. As shown in Figure 3(a), the boosted decision tree regression model shows a higher coefficient of determination than the multiple linear regression model, and reproduces the thermal conductivity of the half-Heusler compounds well. The ML model, such as a simple multiple linear regression is not suitable for ML of the lattice thermal conductivities. This result suggests that the lattice thermal conductivity exhibits a high accuracy owing to the complex interaction of various descriptors. Figure 3(b) shows the feature importance scores for the 55 parameters in ML of boosted decision tree regression. Among the 55 parameters, the top 4 parameters make a significant contribution to the accuracy of ML. It is known that when many 55 parameters are used in ML, the prediction accuracy is reduced because of over-fitting. It is necessary to find the best combination of parameters to improve the prediction accuracy of the lattice thermal conductivity. Figure 3(c) shows the results of the evaluation by the Wrapper Method using backward feature elimination. The backward feature elimination calculates the permutation feature importance of each parameter and sequentially removes the unimportant parameter to find the optimal combination of parameters. The $R^2$ for the ML of thermal conductivity using the top four parameters is the highest $R^2$ of 0.84, which is an improvement over the $R^2$ when all the parameters are considered. The best parameter combination of the important features score for the prediction of the lattice thermal conductivity are in the following order:

(Top 1) $x_{55}$: lattice parameter.

(Top 2) $x_{42}$: the difference between the mean atomic radius of the constituent elements and the atomic radius of the 4*c* site, $(r_1 + r_2 + r_3) / 3 - r_1$.

(Top 3) $x_{33}$: the difference between the mean atomic mass of the constituent elements and the atomic mass of the 4*c* site, $(m_1 + m_2 + m_3) / 3 - m_1$.

(Top 4) $x_{29}$: the sum of the atomic masses, $m_1 + m_2 + m_3$.

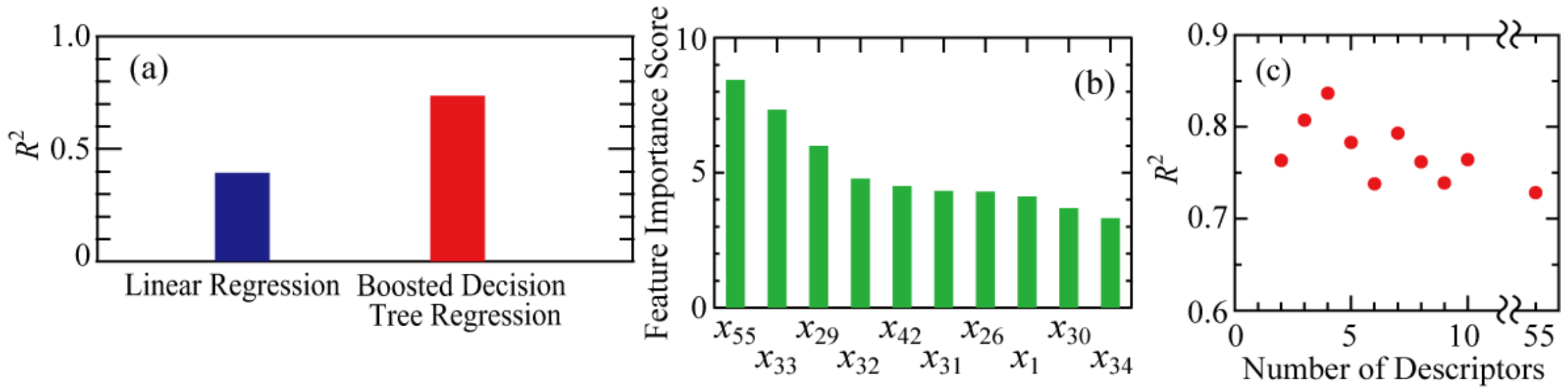

FIG. 3.

(a) $R^2$ for the ML model of multiple linear regression and boosted decision tree regression performed on the calculated lattice thermal conductivity. (b) Top 10 parameters of permutation feature importance for each parameter in the ML model of boosted decision tree regression. (c) Dependence of the number of features on the $R^2$ evaluated by the wrapper Method with backward feature elimination.

Figures 4(a1) and 4(a2) show the results of lattice thermal conductivity predicted by ML for the boosted decision tree regression described with 55 and 4 parameters, respectively. The predictions of thermal conductivity with 4 parameters are in a better agreement with the results of ML with 55 parameters. Figure 4(b) shows the number of deviations between the predicted and calculated lattice thermal conductivities by the ML model of the boosted decision tree regression with 55 and 4 parameters. When 55 parameters were used, the lattice thermal conductivity was overestimated, with a deviation of approximately 8%. Conversely, when 4 parameters were used to describe the lattice thermal conductivity, the accuracy improved to approximately ± 4%. Jesus Carrete *et al* [33] and Jianghui Liu *et al* [34] have previously reported that the accuracy of the lattice thermal conductivity prediction in half-Heusler compounds predicted by ML is approximately ±10%. In this study, our ML model using the lattice parameter as a descriptor and selecting an appropriate combination of atomic radii and atomic masses as a descriptor led to a significant improvement in the accuracy of the prediction of lattice thermal conductivity.

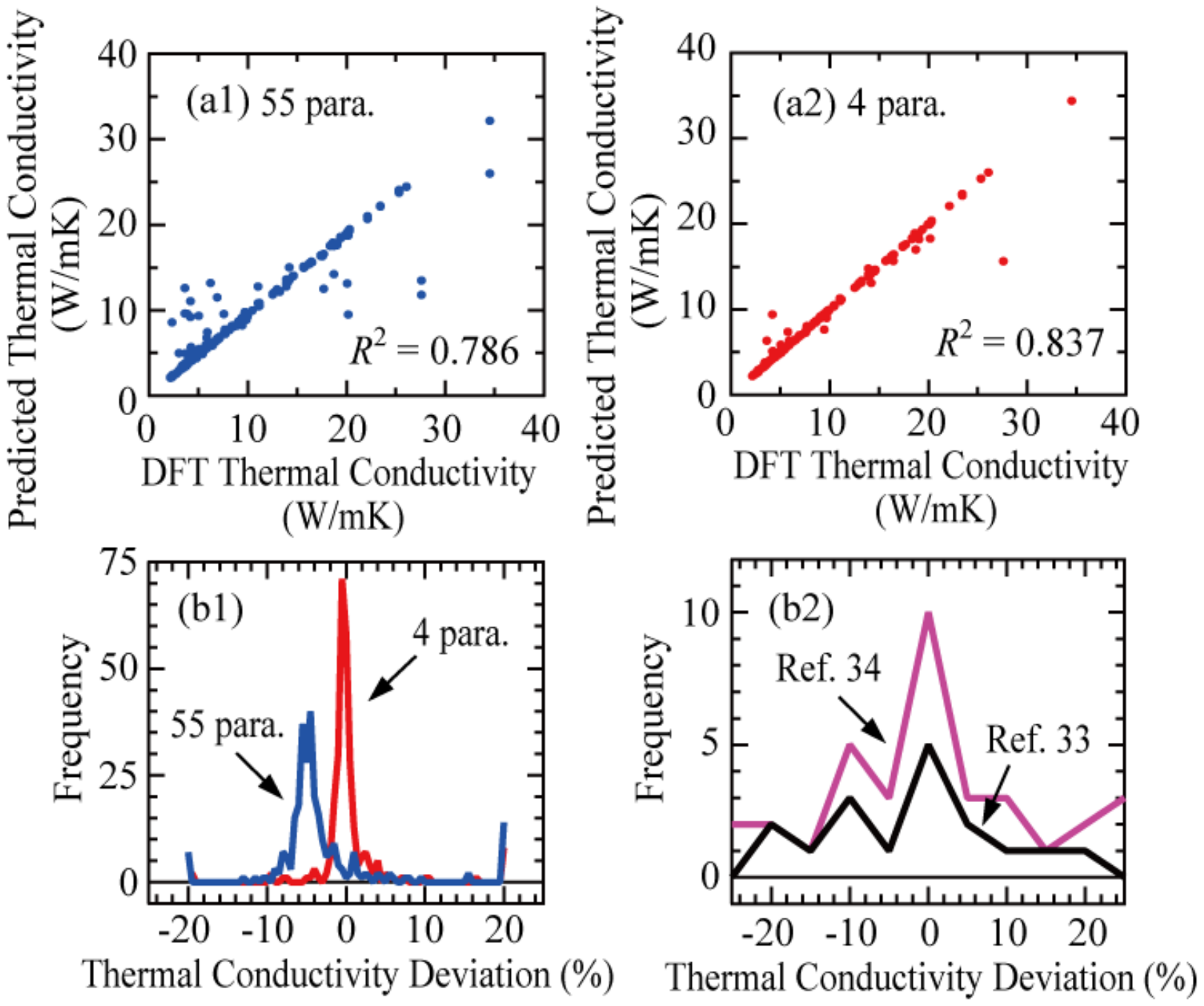

FIG. 4

(a1, (a2) Comparison between the lattice thermal conductivity predicted by the ML model of boosted decision tree regression using 55, and 4 parameters and the lattice thermal conductivity calculated by the DFT calculation. (b1) Number of deviations between the predicted and calculated lattice thermal conductivity using 55 and 4 parameters. (b2) Number of deviations between the predicted and calculated lattice thermal conductivity reported by Jesus Carrete *et al* [33] and Jianghui Liu *et al* [34].

Figure 5 plots the relationship between three parameters and thermal conductivity, which are important in determining the lattice thermal conductivity of half-Heusler compounds. For many compounds with large lattice parameters, the lattice thermal conductivity is below 6, which is a low lattice thermal conductivity material. The lattice thermal conductivity of a solid material is expressed by the equation $\kappa \sim C v l$, where $C$, $v$, and $l$ represent the specific heat, sound velocity, and phonon mean free path, respectively. It is essential to reduce the thermal conductivity to improve the performance of thermoelectric conversion materials. For this purpose, $C$, $v$, and $l$ should be reduced. However, it is difficult to change $C$ and $l$ significantly in the same crystal system. The lattice thermal conductivity is lower in a compound with a large sum of atomic mass, including the structure and small lattice parameter because the sound velocity is expressed as a function inversely proportional to the density. Therefore, for compounds with small lattice parameters, the decrease in thermal conductivity can be qualitatively explained by the decrease in sound velocity. In half-Heusler compounds with lattice parameters of 0.7 to 0.8, the thermal conductivity is below 4 for the systems BaNaSb, KBaSb, CaCdSn, and KSrSb with large differences between the average atomic mass and 4*c* sites in the compounds. Even in half-Heusler compounds with large lattice parameters, small lattice thermal conductivities can be achieved by selecting the heavy elements to occupy the 4*c* site.

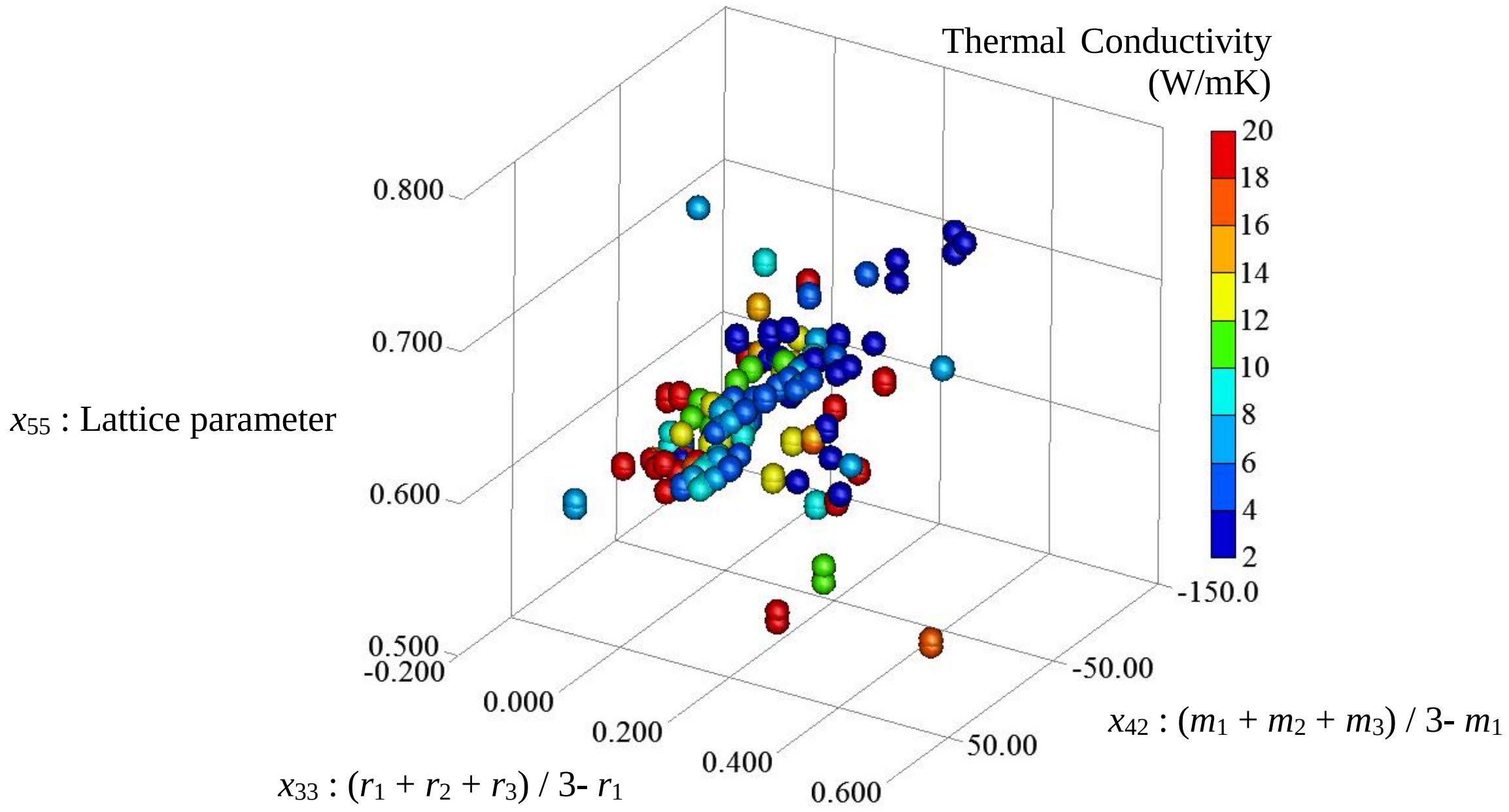

FIG. 5.
3D plots of the three parameters ($x_{33}$, $x_{42}$, $x_{55}$) and predicted lattice thermal conductivity. The magnitude of the lattice thermal conductivity in 3D plotting is shown in color scale.

The material design of half-Heusler compounds has focused on compounds with elements from groups 1 to 4 applied to the 4*a* site, elements from groups 12 to 16 applied to the 4*b* site, and elements from groups 8 to 11 applied to the 4*c* site. This study revealed that many half-Heusler compounds contain groups 14 to 16 elements at the 4*c* site and groups 1 and 2 elements at the 4*a* and 4*b* sites. When heavier elements, such as Sn and Sb, occupy the 4*c* site, the difference between the average atomic weights is particularly large, and such compounds are predicted to show considerably decreased thermal conductivities. In the future, the synthesis of these groups of materials will lead to new low thermal conductivity half-Heusler compounds.

## IV. Conclusion

In this study, we investigated whether the lattice thermal conductivity of half-Heusler compounds can be predicted by ML from the atomic radii and masses of the constituent elements. The results show that the lattice thermal conductivity of the half-Heusler compounds can be predicted with an accuracy of ±4% using the predicted lattice parameters and the atomic radii and masses. In addition to the conventional material design in which a material with a small lattice parameter and low density has a low thermal conductivity, it was found that even for materials with a large lattice parameter, one can design a material with a low thermal conductivity by selecting elements occupying the 4*c* site of the half-Heusler structure with a larger atomic mass than those occupying the 4*a* and 4*b* sites. Using the results of ML, the thermal conductivity of unknown half-Heusler compounds can be instantly predicted. It is expected that the search for half-Heusler compounds with the desired thermal conductivity will become easier and the development of functional materials in a short time and at a low cost will be advanced in the future.

## SUPPLEMENTARY MATERIALS

See supplementary materials for the periodic table of the site selection system of the 143 half-Heusler compounds and the complete results of the calculated lattice parameter and thermal conductivity and list of parameters for ML used for thermal conductivity prediction

## ACKNOWLEDGEMENTS

The computations were performed by the Research Center for Computational Science, Okazaki, Japan. This study was partly supported by the Japan Society for the Promotion of Science's (JSPS) Grants-in-Aid for Scientific Research (KAKENHI) (C) (17K06771, 18K04700, 18K04748, 20K05100, and 20K05060). We would like to thank Editage (www.editage.com) for English language editing.

## SUPPLEMENTARY MATERIALS

Figure S1 shows the periodic table of the site selection system of the constituent elements for the 143 half-Heusler compounds. The elements from groups 8 to 11 occupy almost 4*c* sites; groups 1 to 6 and the lanthanide and actinide series occupy 4*a* sites and some elements also occupy 4*b* sites. The elements of groups 12–16 occupy 4*b* sites as well as 4*c* sites for most of the elements. This suggests that the site selection in half-Heusler compounds is not determined solely by the relationship between the atomic radii of the elements at each site but also by the atomic masses at each site, which is a complex balance of the two elemental information.

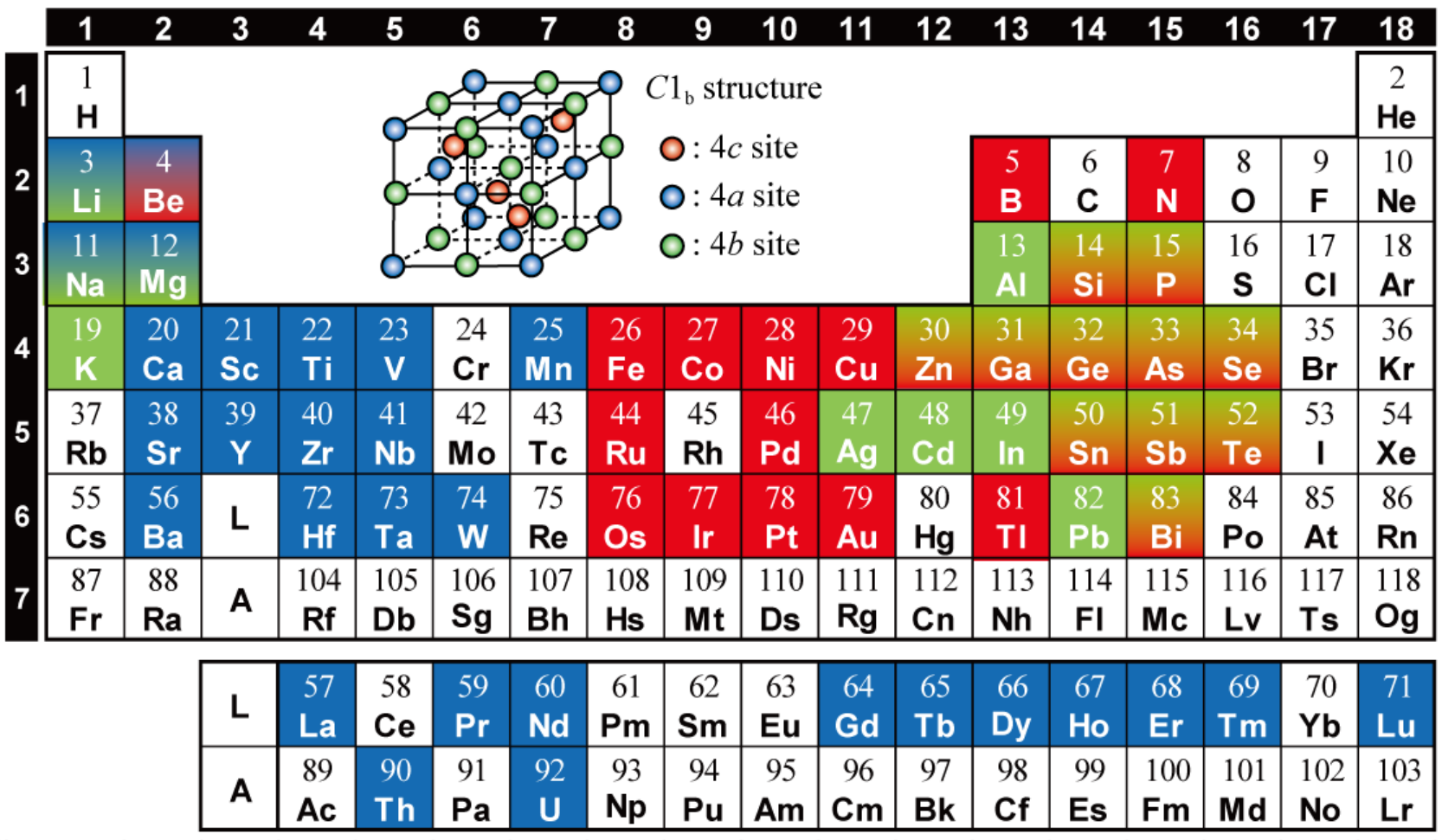

FIG. S1

Periodic table summarizing the site-selective system for each element in the half-Heusler compound determined by DFT calculations.

TABLE S1

Table of the Materials ID, Chemcal formula, optimal site locations, lattice parameters, and thermal conductivities for 143 half-Heusler compounds. The atoms occupying each site and the corresponding lattice parameter are shown for the optimal site selection with the lowest total energy. Thermal conductivities are calculated using the lattice thermal conductivity calculation code Phonopy3 for the best site combination.

| Materials ID | Chemical Fomula | 4*c* site | 4a site | 4b site | L. P. | T. C. |
|---|---|---|---|---|---|---|
| 2894 | ScSnAu | Au | Sc | Sn | 0.653 | 7.09 |
| 3161 | LiAlSi | Si | Li | Al | 0.594 | 5.90 |
| 3432 | ScNiSb | Ni | Sc | Sb | 0.612 | 13.91 |
| 3462 | TmSnAu | Au | Tm | Sn | 0.669 | 4.61 |
| 3522 | MgCuSb | Cu | Mg | Sb | 0.625 | 3.68 |
| 3716 | TbNiSb | Ni | Tb | Sb | 0.639 | 9.06 |
| 4025 | TmNiSb | Ni | Tm | Sb | 0.631 | 9.85 |
| 4174 | HoNiSb | Ni | Ho | Sb | 0.635 | 9.69 |
| 4262 | BeAlB | B | Be | Al | 0.496 | 18.71 |
| 4510 | DyNiSb | Ni | Dy | Sb | 0.636 | 9.46 |
| 4964 | YSbPt | Pt | Y | Sb | 0.663 | 3.23 |
| 5177 | LuSnAu | Au | Lu | Sn | 0.666 | 4.72 |
| 5640 | ErSnAu | Au | Er | Sn | 0.671 | 4.53 |
| 5920 | LiAlGe | Ge | Li | Al | 0.601 | 5.08 |
| 5967 | TiCoSb | Co | Ti | Sb | 0.590 | 22.13 |
| 7173 | ScSbPt | Pt | Sc | Sb | 0.640 | 7.67 |
| 7575 | LiZnN | N | Li | Zn | 0.493 | 17.39 |
| 9124 | LiZnAs | As | Li | Zn | 0.598 | 13.89 |
| 9437 | NbFeSb | Fe | Nb | Sb | 0.596 | 23.43 |
| 10194 | LuSbPt | Pt | Lu | Sb | 0.655 | 6.42 |
| 10687 | LiCdP | P | Li | Cd | 0.614 | 20.07 |
| 11242 | DyPbAu | Au | Dy | Pb | 0.685 | 3.35 |
| 11520 | YNiSb | Ni | Y | Sb | 0.638 | 9.03 |
| 11836 | ErSbPd | Pd | Er | Sb | 0.657 | 7.65 |
| 11839 | GdSbPt | Pt | Gd | Sb | 0.666 | 3.34 |
| 11869 | HfSnPd | Pd | Hf | Sn | 0.637 | 13.99 |
| 12558 | LiMgAs | As | Li | Mg | 0.620 | 20.18 |
| 13308 | HoGeAu | Au | Ho | Ge | 0.653 | 4.34 |
| 16313 | TbSbPt | Pt | Tb | Sb | 0.664 | 3.82 |
| 16314 | TmSbPt | Pt | Tm | Sb | 0.657 | 5.86 |
| 16327 | DySbPt | Pt | Dy | Sb | 0.662 | 4.43 |
| 16329 | ErSbPt | Pt | Er | Sb | 0.659 | 5.41 |
| 16376 | HoSbPt | Pt | Ho | Sb | 0.661 | 4.95 |
| 19886 | ThSnPt | Pt | Th | Sn | 0.683 | 4.15 |
| 20185 | LuNiSb | Ni | Lu | Sb | 0.628 | 9.97 |
| 20269 | MnGaPt | Pt | Mn | Ga | 0.600 | 2.33 |
| 20415 | GdPbAu | Au | Gd | Pb | 0.689 | 3.18 |
| 20514 | GdSnAu | Au | Gd | Sn | 0.679 | 3.73 |
| 20952 | TiNiSb | Ni | Ti | Sb | 0.596 | 3.64 |
| 21272 | ErNiSb | Ni | Er | Sb | 0.633 | 9.68 |
| 21425 | UNiSn | Ni | U | Sn | 0.636 | 5.36 |
| 22786 | ThNiSn | Ni | Th | Sn | 0.660 | 7.15 |
| 30377 | ErPbAu | Au | Er | Pb | 0.682 | 3.51 |
| 30389 | HoPbAu | Au | Ho | Pb | 0.684 | 3.43 |
| 30390 | HoSnAu | Au | Ho | Sn | 0.673 | 4.46 |
| 30413 | TbPbAu | Au | Tb | Pb | 0.687 | 3.26 |
| 30453 | DyBiPt | Pt | Dy | Bi | 0.675 | 6.23 |
| 30454 | ErBiPt | Pt | Er | Bi | 0.672 | 6.35 |
| 30455 | HoBiPt | Pt | Ho | Bi | 0.674 | 6.46 |
| 30457 | LuNiBi | Ni | Lu | Bi | 0.641 | 9.83 |
| 30459 | ScNiBi | Ni | Sc | Bi | 0.627 | 9.06 |
| 30460 | YNiBi | Ni | Y | Bi | 0.651 | 5.91 |
| 30847 | TiSnPt | Pt | Ti | Sn | 0.624 | 11.18 |
| 30848 | USnPt | Pt | U | Sn | 0.667 | 5.41 |
| 31451 | ZrCoBi | Co | Zr | Bi | 0.624 | 17.68 |
| 31454 | TaSbRu | Ru | Ta | Sb | 0.620 | 16.23 |
| 31455 | VSbRu | Ru | V | Sb | 0.605 | 6.90 |
| 31457 | ZrSbRu | Ru | Zr | Sb | 0.636 | 3.36 |
| 36111 | LiMgP | P | Li | Mg | 0.600 | 19.94 |
| 505297 | NbSbRu | Ru | Nb | Sb | 0.620 | 15.65 |
| 567418 | HoSbPd | Pd | Ho | Sb | 0.659 | 7.72 |
| 567422 | GdNiBi | Ni | Gd | Bi | 0.653 | 8.12 |
| 567636 | VFeSb | Fe | V | Sb | 0.580 | 14.20 |
| 568269 | TmNiBi | Ni | Tm | Bi | 0.644 | 9.36 |
| 569197 | GdNiSb | Ni | Gd | Sb | 0.641 | 8.38 |
| 569779 | ScSbPd | Pd | Sc | Sb | 0.639 | 11.07 |
| 570213 | LiMgBi | Bi | Li | Mg | 0.682 | 15.77 |
| 620271 | GdBiPt | Pt | Gd | Bi | 0.679 | 5.54 |
| 621592 | YPbAu | Au | Y | Pb | 0.687 | 3.28 |
| 924128 | HfNiSn | Ni | Hf | Sn | 0.612 | 16.44 |
| 924129 | ZrNiSn | Ni | Zr | Sn | 0.616 | 19.05 |
| 924130 | TiNiSn | Ni | Ti | Sn | 0.596 | 17.54 |
| 961646 | TiTeOs | Os | Ti | Te | 0.618 | 12.96 |
| 961649 | ZrFeTe | Fe | Zr | Te | 0.610 | 20.00 |
| 961653 | FeSiW | Fe | Si | W | 0.556 | 6.21 |
| 961657 | YNiP | Ni | Y | P | 0.598 | 2.37 |
| 961659 | TiSiPd | Pd | Ti | Si | 0.589 | 25.31 |
| 961660 | TiFeSe | Fe | Ti | Se | 0.562 | 27.60 |
| 961661 | ZrSiPd | Pd | Zr | Si | 0.612 | 18.35 |
| 961665 | MgScGa | Ga | Mg | Sc | 0.647 | 3.88 |
| 961673 | TiFeTe | Fe | Ti | Te | 0.589 | 20.30 |
| 961675 | ScNiP | Ni | Sc | P | 0.569 | 8.87 |
| 961678 | ScCoTe | Co | Sc | Te | 0.607 | 12.70 |
| 961682 | TiSnPd | Pd | Ti | Sn | 0.622 | 11.03 |
| 961684 | LiCaAs | As | Li | Ca | 0.667 | 20.16 |
| 961685 | NaCaAs | As | Na | Ca | 0.696 | 7.57 |
| 961687 | ZrSnPd | Pd | Zr | Sn | 0.641 | 14.04 |
| 961693 | ZrInAu | Au | Zr | In | 0.653 | 9.42 |
| 961697 | ScGeAu | Au | Sc | Ge | 0.630 | 12.52 |
| 961698 | LiZnP | P | Li | Zn | 0.576 | 9.72 |
| 961706 | TiSiPt | Pt | Ti | Si | 0.591 | 34.51 |
| 961711 | ZrSiPt | Pt | Zr | Si | 0.615 | 14.11 |
| 961713 | ZrSnPt | Pt | Zr | Sn | 0.643 | 13.23 |
| 961774 | BaNaSb | Sb | Ba | Na | 0.796 | 2.17 |
| 962063 | LaMgTl | Tl | La | Mg | 0.731 | 6.26 |
| 962068 | CaMgSn | Sn | Ca | Mg | 0.716 | 4.19 |
| 962069 | LiAgTe | Te | Li | Ag | 0.664 | 3.38 |
| 962078 | CaCdSi | Si | Ca | Cd | 0.670 | 3.39 |
| 1008624 | YBiPd | Pd | Y | Bi | 0.674 | 4.24 |
| 1008680 | TiGePt | Pt | Ti | Ge | 0.600 | 19.40 |
| 1008858 | NdBiPd | Pd | Nd | Bi | 0.687 | 4.59 |
| 1009006 | LiCdAs | As | Li | Cd | 0.634 | 16.44 |
| 1009132 | HoBiPd | Pd | Ho | Bi | 0.672 | 6.78 |
| 1009543 | DyBiPd | Pd | Dy | Bi | 0.673 | 6.93 |
| 1018118 | TmSbPd | Pd | Tm | Sb | 0.656 | 7.66 |
| 1018135 | LiCdAs | As | Li | Cd | 0.634 | 16.49 |
| 1018139 | HoNiBi | Ni | Ho | Bi | 0.647 | 9.44 |
| 1076916 | GdBiPd | Pd | Gd | Bi | 0.677 | 6.80 |
| 1093991 | ZrAsIr | Ir | Zr | As | 0.618 | 10.43 |
| 1094088 | NbCoSn | Co | Nb | Sn | 0.597 | 18.62 |
| 1100391 | NaBeSb | Be | Na | Sb | 0.625 | 3.62 |
| 1100392 | LiSiB | B | Li | Si | 0.501 | 11.09 |
| 1100393 | LiAgSe | Se | Li | Ag | 0.628 | 2.72 |
| 1100403 | KBaSb | Sb | K | Ba | 0.828 | 2.88 |
| 1100404 | VAsRu | Ru | V | As | 0.580 | 5.04 |
| 1100409 | LiCaSb | Sb | Li | Ca | 0.712 | 26.07 |
| 1100411 | CaCdSn | Sn | Ca | Cd | 0.712 | 3.02 |
| 1100412 | TiTeRu | Ru | Ti | Te | 0.615 | 14.54 |
| 1100420 | NaCaSb | Sb | Na | Ca | 0.740 | 5.45 |
| 1100424 | LiMgSb | Sb | Li | Mg | 0.667 | 14.63 |
| 1100425 | LiCdSb | Sb | Li | Cd | 0.674 | 3.51 |
| 1100429 | KSrSb | Sb | K | Sr | 0.804 | 3.66 |
| 1100430 | LiZnSb | Zn | Li | Sb | 0.632 | 4.81 |
| 1100433 | CaZnSi | Si | Ca | Zn | 0.648 | 2.46 |
| 1100436 | LiInSi | Si | Li | In | 0.628 | 3.61 |
| 1206667 | PrNiBi | Ni | Pr | Bi | 0.668 | 4.47 |
| 1206679 | YBiPt | Pt | Y | Bi | 0.676 | 4.61 |
| 1206681 | ScBiPd | Pd | Sc | Bi | 0.653 | 5.85 |
| 1206686 | TbNiBi | Ni | Tb | Bi | 0.651 | 8.57 |
| 1206712 | ErNiBi | Ni | Er | Bi | 0.646 | 9.68 |
| 1206717 | LaBiPd | Pd | La | Bi | 0.696 | 4.35 |
| 1206719 | NdNiBi | Ni | Nd | Bi | 0.665 | 5.25 |
| 1206720 | PrBiPd | Pd | Pr | Bi | 0.690 | 4.19 |
| 1206744 | TmBiPd | Pd | Tm | Bi | 0.669 | 6.28 |
| 1206953 | ErBiPd | Pd | Er | Bi | 0.670 | 6.53 |
| 1206989 | TbBiPt | Pt | Tb | Bi | 0.677 | 5.90 |
| 1206992 | LuBiPt | Pt | Lu | Bi | 0.668 | 5.77 |
| 1207057 | TmBiPt | Pt | Tm | Bi | 0.670 | 5.94 |
| 1207082 | LiMgSb | Sb | Li | Mg | 0.667 | 14.63 |
| 1207177 | TbBiPd | Pd | Tb | Bi | 0.675 | 6.93 |
| 1207185 | LuBiPd | Pd | Lu | Bi | 0.666 | 5.75 |
| 1216635 | TiSnIr | Ir | Ti | Sn | 0.622 | 3.90 |
| 1225491 | DySbPd | Pd | Dy | Sb | 0.661 | 7.67 |
| | | | | | | |

TABLE S2

List of parameters for machine learning used for thermal conductivity prediction. It is listed by various combinations of atomic mass ($m_1$, $m_2$, $m_3$) and atomic radius ($r_1$, $r_2$, $r_3$) of elementals at 4*c*, 4*a*, 4*b* site. The lattice parameter, $x_{55}$, uses the predicted values of the lattice parameters obtained by the multiple linear regression in Fig. 2(b2).

| | Parameters |
|---|---|
| $x_1$ | $m_1$ |
| $x_2$ | $m_2$ |
| $x_3$ | $m_3$ |
| $x_4$ | $r_1$ |
| $x_5$ | $r_2$ |
| $x_6$ | $r_3$ |
| $x_7$ | ${m_1}^2$ |
| $x_8$ | ${m_2}^2$ |
| $x_9$ | ${m_3}^2$ |
| $x_{10}$ | ${r_1}^2$ |
| $x_{11}$ | ${r_2}^2$ |
| $x_{12}$ | ${r_3}^2$ |
| $x_{13}$ | ${m_1}^3$ |
| $x_{14}$ | ${m_2}^3$ |
| $x_{15}$ | ${m_3}^3$ |
| $x_{16}$ | ${r_1}^3$ |
| $x_{17}$ | ${r_2}^3$ |
| $x_{18}$ | ${r_3}^3$ |
| $x_{19}$ | $\sqrt{m_1}$ |

| | Parameters |
|---|---|
| $x_{20}$ | $\sqrt{m_2}$ |
| $x_{21}$ | $\sqrt{m_3}$ |
| $x_{22}$ | $\sqrt{r_1}$ |
| $x_{23}$ | $\sqrt{r_2}$ |
| $x_{24}$ | $\sqrt{r_3}$ |
| $x_{25}$ | $m_2/m_1$ |
| $x_{26}$ | $r_3/m_1$ |
| $x_{27}$ | $r_2/r_1$ |
| $x_{28}$ | $r_3/r_1$ |
| $x_{29}$ | $m_1 + m_2 + m_3$ |
| $x_{30}$ | $r_1 + r_2 + r_3$ |
| $x_{31}$ | $\left(({m_1}^2 + {m_2}^2 + {m_3}^2)/3\right)^2$ |
| $x_{32}$ | $\left(({r_1}^2 + {r_2}^2 + {r_3}^2)/3\right)^2$ |
| $x_{33}$ | $x_{29}/3 - m_1$ |
| $x_{34}$ | $x_{29}/3 - m_2$ |
| $x_{35}$ | $x_{29}/3 - m_3$ |
| $x_{36}$ | $\lvert x_{33} \rvert$ |
| $x_{37}$ | $\lvert x_{34} \rvert$ |
| $x_{38}$ | $\lvert x_{35} \rvert$ |

| | Parameters |
|---|---|
| $x_{39}$ | ${x_{33}}^2$ |
| $x_{40}$ | ${x_{34}}^2$ |
| $x_{41}$ | ${x_{35}}^2$ |
| $x_{42}$ | $x_{30}/3 - r_1$ |
| $x_{43}$ | $x_{30}/3 - r_2$ |
| $x_{44}$ | $x_{30}/3 - r_3$ |
| $x_{45}$ | $\lvert x_{42} \rvert$ |
| $x_{46}$ | $\lvert x_{43} \rvert$ |
| $x_{47}$ | $\lvert x_{44} \rvert$ |
| $x_{48}$ | ${x_{42}}^2$ |
| $x_{49}$ | ${x_{43}}^2$ |
| $x_{50}$ | ${x_{44}}^2$ |
| $x_{51}$ | ${r_1}^2 + {r_2}^2$ |
| $x_{52}$ | ${r_1}^2 + {r_3}^2$ |
| $x_{53}$ | $\sqrt{x_{51}}$ |
| $x_{54}$ | $\sqrt{x_{52}}$ |
| $x_{55}$ | Predicted lattice parameter |